# Graph Neural Network-based Power Flow Model

Mingjian Tuo
*Student Member, IEEE*
Department of Electrical and Computer Engineering
University of Houston
Houston, TX, USA
mtuo@uh.edu

Xingpeng Li
*Senior Member, IEEE*
Department of Electrical and Computer Engineering
University of Houston
Houston, TX, USA
xli82@uh.edu

Tianxia Zhao

Industrial AI Team Lead
Shell
Houston, TX, USA
Tina.T.Zhao@shell.com

*Abstract*—Power flow analysis plays a crucial role in examining the electricity flow within a power system network. By performing power flow calculations, the system's steady-state variables, including voltage magnitude, phase angle at each bus, and active/reactive power flow across branches, can be determined. While the widely used DC power flow model offers speed and robustness, it may yield inaccurate line flow results for certain transmission lines. This issue becomes more critical when dealing with renewable energy sources such as wind farms, which are often located far from the main grid. Obtaining precise line flow results for these critical lines is vital for next operations. To address these challenges, data-driven approaches leverage historical grid profiles. In this paper, a graph neural network (GNN) model is trained using historical power system data to predict power flow outcomes. The GNN model enables rapid estimation of line flows. A comprehensive performance analysis is conducted, comparing the proposed GNN-based power flow model with the traditional DC power flow model, as well as deep neural network (DNN) and convolutional neural network (CNN). The results on test systems demonstrate that the proposed GNN-based power flow model provides more accurate solutions with high efficiency comparing to benchmark models.

## I. INTRODUCTION

Power flow is necessary to analyze the steady-state of the system, it involves studying the flow of electrical power within a transmission network and determining various power system operations. Current AC power flow (ACPF) problems are usually solved using iterative methods such as Gauss-Seidel (GS) method and Newton-Raphson (NR). However, for a large-scale system, it is not practical to use AC power flow equation-based methods for decision making or fast screening of the system [1]. The decarbonization of the electricity generation relies on the integration of converter-based renewable energy sources (RES) over the past decades. RES such as solar and wind power exhibit variability and intermittency due to factors like weather conditions and time of day. This variability introduces fluctuations in generation, which can affect power flow patterns. Applying ML to solve these challenging situations is critical for development of clean and green energy of the future and it could help achieve a reasonable statistical evaluation of the risk. Applications of machine learning (ML) toward renewable energy have been widely researched and studied in recent years [1]-[4]. Alternatively, the constraints and data associated with security constrained unit commitment are studied using ML model [5]. ML as an advanced algorithm to predict generation of renewable sources has been proposed in [6]. In [7], deep reinforcement learning was investigated as a possible control strategy for power systems with multiple renewable energy sources.

Compared to traditional computational approaches, machine learning algorithms have an intrinsic generalization capability with greater computational efficiency and scalability [8]. Since machine learning algorithms have the ability to learn complex nonlinear input-output relationships, reflexively adapting themselves to the data, they have been used to predict voltage magnitude and phase angle at each bus [9]. Similarly, predictions of initial system variables based on ML mode is implemented to decrease the solution iterations and time for NR based ACPF model in [10]. However, the spatial information embedded in the predictions for active power flow for each branch were not considered.

Power system is an interconnected network of generators and loads which has embedded graphical information. The graph structure of the power system consists of nodes (buses) and edges (branches) [11]. The branches in the power system are undirected, such graphs provide information on buses and their connections. Graph neural network (GNN) is a class of artificial neural network (NN) that has advantages in processing graphical data which have explicit topological graph correlation embedded in graph structures such as power system [12]. They were first introduced by Scarselli et al. [13] and further developed in Li et al. [14]. The key idea of GNN is to iteratively propagate messages through the edges of the graph structure. Ref. [15] proposes a purely data-driven approach based on graph convolutional networks and reports promising results on real-world power grids such as Texas and East Coast systems. An encoder/decoder framework is employed in [16] and a messages propagation mechanism among neighboring nodes to solve PF. However, these methods have strong physical assumption, dependency on preprocessing [17]. Moreover, a thorough analysis regarding the performance of the NN model vs. the non-iterative DC power flow (DCPF) method was not conducted; and the predictions for active power flow for each branch were not considered.

To bridge the aforementioned gaps, a GNN-based power flow prediction model is proposed, the detailed research was performed regarding model selection and how to maximize the performance of the proposed GNN model. The model was trained and tested against multiple systems to assess how accurate it could predict the outputs. Comparing to DC power flow model, its performance and effectiveness are evaluated and demonstrated.

The remainder of this paper is organized as follows. Section II discusses the power system power flow calculation and overview of GNN methodology. Section III presents the construction of proposed GNN model. Section IV shows the simulation results and evaluates the performance of the proposed GNN method. Section V concludes this paper and presents future work.

## II. PRELIMINARIES

### A. Power Flow Calculation

Power flow calculation is a computational method used to determine the steady-state operating conditions of a power system. It involves solving a set of power flow equations to determine voltage magnitudes, phase angles, and power flows throughout the system. The mathematical model of the power system network the buses (nodes), transmission lines (branches), generators, loads, and other components. The network model should capture the electrical characteristics and connectivity of the system accurately. The nodal power balance equations are listed as follows [18],

$$P_i - V_i \sum_{j \in \{i,1,\ldots,N\}} V_j(G_{ij} \cos \theta_{ij} + B_{ij} \sin \theta_{ij}) = 0, \quad (1)$$

$$Q_i - V_i \sum_{j \in \{i,1,\ldots,N\}} V_j(G_{ij} \sin \theta_{ij} + B_{ij} \cos \theta_{ij}) = 0, \quad (2)$$

where, $P_i$ and $Q_i$ are the active and reactive power injections at each node respectively. The summation terms represent the active or reactive power injections or withdrawals from the electrical network at a given node $i$. $V_i$ and $V_j$ are the voltage magnitude for the two end buses of a transmission line. $G_{ij}$ and $B_{ij}$ are the corresponding conductance and susceptance of a branch. The phase angle $\theta_{ij}$ is the difference in voltage phase angles of the two end buses of a branch.

An accurate and comprehensive power flow analysis method is achieved by employing conventional ACPF techniques like the NR method and GS method. This approach takes into account the non-linear properties of power system components, including transformers and transmission lines However, ACPF solves a set of nonlinear equations that represent the power flow equations based on Kirchhoff's laws and other system constraints by iteratively calculation. Due to the inherent and complex iterative nature of these algorithms, it may suffer divergency and may not be beneficial for some online monitoring applications or to be integrated into optimization based scheduling and dispatching models. A non-iterative method called DC power flow can be used for fast power flow solution [19]. DCPF represents the network as an equivalent DC network, reducing the complexity of calculations. It simplifies the power flow equations by assuming that voltage magnitudes remain constant, and the phase angles are small. With DCPF, the solutions to steady-state active power flow can be found very quickly. However, approximation error introduced by the assumptions made earlier may lead to inaccurate results. Thus, it is desirable to develop a new method that can provide fast and accurate power flow solutions.

### B. Machine Learning Overview

Machine learning is a computer algorithm that can automatically improve/learn through experience by using historical data. Given a system with $N$ generators, the model will be built using historical dataset to make predictions,

$$\hat{h}^{PF}(s_t, u_t, d_t, r_t) = PF, \quad (3)$$

where $\hat{h}^{PF}$ is the nonlinear power flow prediction model; $s_t$ denotes the system states, and $u_t$ is the generation dispatch at period $t$. $d_t$ and $r_t$ denote the load profile and renewable forecast, respectively. The overview of the power flow model is shown in Fig. 1.

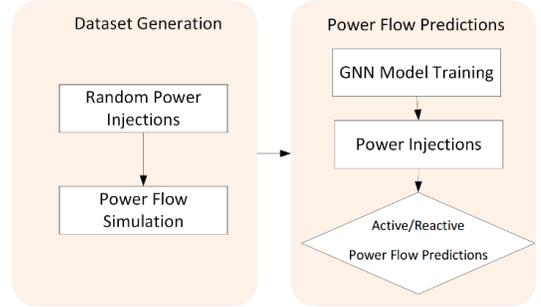

Fig. 1. Overview of power flow model.

For a basic power flow model, training data (injections) are multiplied with weight vector $W$. Then, the results are added by bias $b$ and mapped to an output value after applying an activation function. The choice of activation functions can vary depending on the model selected. Throughout the training process, the weight vector undergoes continuous updates until the error falls below a predetermined threshold or a specified number of epochs is reached [20].

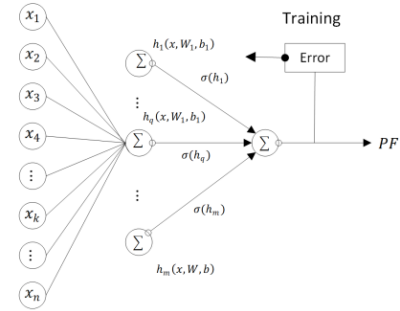

Fig. 2. Example of power flow model training process.

In order to handle the mapping with high nonlinearity, a power flow model typically involves multiple nodes (artificial neurons) with several hidden layers as shown in Fig. 3. The connections between nodes (vector weights) reflect the signal strength between each neuron. A neural network consisting of multiple layers performs distinct transformations on input data. Each layer or node within the network learns specific features from the input. By leveraging an NN-based power flow model, complex input-output relationships can be learned, which may be challenging to grasp or program using conventional algorithms.

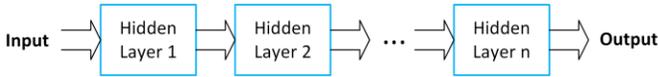

Fig. 3. Neural network with multiple layers.

## III. GRAPH NEURAL NETWORK BASED POWER FLOW MODEL

CNNs have inherent limitations when it comes to handling graphical data that contains explicit topological graph correlations [12]. However, recent progress in CNNs has led to the resurgence of GNNs, which are neural networks specialized in processing and extracting knowledge from graph-structured data. A power system can be viewed as a graph comprising nodes (buses) and edges (branches) that denote the connections between nodes. GNNs are specifically designed to capture the intricate dependencies and relationships present in graph data. To achieve this, GNNs have been developed by extending the convolution operation to graphs and, more generally, to non-Euclidean spaces. Previous studies in [21] have proved that GNN provides state-of-arts performance in graph analysis tasks.

The input vector $X$ concatenates information about the electrical power that is being produced and consumed everywhere on the grid. The branches in the power system are undirected, such graphs provide information on buses and their connections. Specifically, each generation $g \in G$ is defined by an active power infeed $P_g$ (in MW) and a reactive power $Q_g$ (in MVar). Therefore, each generation is defined by a 2-dimensional information. Similarly, a nodal load $d \in D$ is defined by an active power consumption $P_d$ (in MW) and a reactive power consumption $Q_d$ (in MVar). Thus, the injection vector $X$ is a vector that concatenates all these injection characteristics including the initial voltage magnitude on each bus.

The convolution operator in propagation module is used to aggregate information from neighbors. Considering $\mathcal{G} = (\mathcal{V}, \mathcal{E})$ as an undirected graph representing a power system, where $\mathcal{V} \in \mathbb{R}^N$ denotes its nodes and $\mathcal{E} \in \mathbb{R}^K$ denotes its edges. Let $A \in \mathbb{R}^{N \times N}$ be the adjacency matrix of $\mathcal{G}$, we can define a renormalization equation as,

$$V = \widetilde{D}^{-\frac{1}{2}} \widetilde{A} \widetilde{D}^{-\frac{1}{2}}, \quad (4)$$

where $\widetilde{A} = A + I_N$ represents an adjacency matrix with added self-connections, and $I_N$ is the identity matrix. The adjacency matrix encodes the way injections are connected to edges. Typically, the element at $(i, j)$ of the adjacency matric $A$ is defined as follows,

$$A_{ij} = \begin{cases} 1; & \text{if } \mathcal{V}_i, \mathcal{V}_j \in \mathcal{V}, (\mathcal{V}_i, \mathcal{V}_j) \in \mathcal{E} \\ 0; & \text{if } \mathcal{V}_i, \mathcal{V}_j \in \mathcal{V}, (\mathcal{V}_i, \mathcal{V}_j) \notin \mathcal{E} \end{cases}, \quad (5)$$

where $(\mathcal{V}_i, \mathcal{V}_j)$ denotes the branches from $i$ to $j$. The diagonal degree matrix $\widetilde{D}$ for $\mathcal{G}$ is defined as $\widetilde{D}_{ii} = \sum_j \widetilde{A}_{ij}$.

GNNs update the representations of nodes in a graph by aggregating information from their neighboring nodes. This process involves iterative message-passing steps, where each node incorporates both local and global information to update its representation. By doing so, GNNs capture both the local structure and the broader context of the graph. The graph convolutional activation is defined as follows,

$$F^l(X, A) = \sigma(V F^{(l-1)}(X, A) W_k^l + b^l), \quad (6)$$

where $F^l$ is the convolutional activations, $W_k^l$ and $b^l$ are the trainable convolutional weights matrix and bias matrix at the $l$-th layer; $F^0 = X$ is the input matrix. In this step, we iteratively update the latent state of each of the n power lines by performing latent leaps that depend on the value of their direct neighbors.

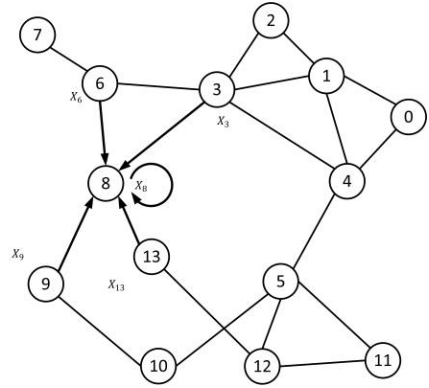

Fig. 4. Example of message passing mechanism in GNN of IEEE 14-bus system.

Fig. 4 demonstrates the message passing mechanism in forward propagation, a target node (bus 8) receiving information from its neighboring nodes. The output we want to predict is the flows through the lines at every line.

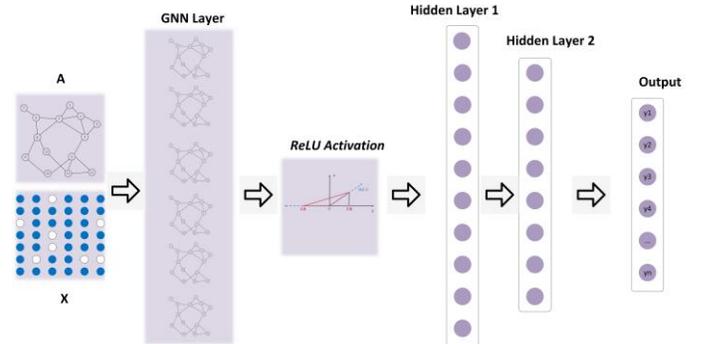

Fig. 5. Illustration of the proposed GNN neural network.

The proposed power flow model as shown in Fig. 5 has one GNN layer for embedding step, two hidden layers are considered as decoding step. The Rectified Linear Unit (ReLU) is chosen as the activation function for forward propagation. It allows a small, non-zero constant gradient to pass through to minimize the vanishing gradient problem during the training phase the model. The decoding step decodes the embedded data to the output space which is the steady-state voltage magnitude for each bus and the power flow on each branch.

Mean square error (MSE) loss was used to measure the performance of the model during the training process. MSE measures the average squared difference between actual and predicted outputs. The MSE loss function used in this paper is defined in (7),

$$MSE = \frac{1}{N} \sum_{i=1}^{N} (y_i - \tilde{y}_i)^2, \quad (7)$$

where $y_i$ denotes the actual output value while $\tilde{y}_i$ denotes the GNN model estimated output value, and $N$ denotes the number of sample points.

## IV. CASE STUDIES

The proposed GNN model was trained to predict voltage magnitude and active power flow on multiple systems with different sizes: the IEEE 14-bus, and IEEE 24-bus test systems. For data generation process, each load and generator dispatch settings are randomly perturbed between [85%, 115%] with uniform distribution, where the load data in the case files is applied as the base value. For active power control corresponding to the variation on total system load values. The power flow data generation is operated in Python 3.8 using pypower. The GNN-based power flow model is trained using pytorch on NVIDIA RTX 2070 GPUs. The generated data set was divided into three groups: 80% for training, 20% for validation.

The proposed GNN model is compared with DNN and CNN model, The following metrics are used to demonstrate the prediction accuracy: (1) maximum error (MAX-E), (2) median absolute error (MED-E), (3) mean absolute percentage error (MAPE), and (4) R2 score, which is defined as follows.

$$R^2(y_i, \hat{y}_i) = 1 - \frac{\sum_{i=1}^{N}(y_i - \tilde{y}_i)^2}{\sum_{i=1}^{N}(y_i - \bar{y})^2} \qquad (8)$$

where $\bar{y}$ is the mean of actual labels. $R^2$ score gives some information about the goodness of model fitting. In regression, the $R^2$ coefficient of determination is a statistical measure of how well the regression predictions approximate the real data points.

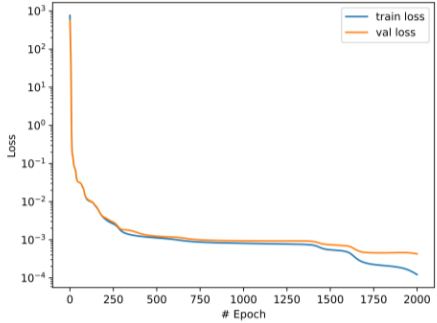

Fig. 6. Training curve of GNN neural network on 14-bus system.

Table I
Prediction Accuracy of Power Flow with Different Tolerance Comparing Different Models

| System | Tolerance | 1% | 2% | 3% | 4% | 5% |
|---|---|---|---|---|---|---|
| 14 Bus | DNN | 66.65 | 76.41 | 82.54 | 86.48 | 88.92 |
|  | CNN | 69.60 | 79.86 | 86.00 | 89.14 | 91.07 |
|  | GNN | 89.60 | 94.24 | 95.59 | 96.26 | 96.66 |
| 24 Bus | DNN | 71.90 | 87.95 | 91.48 | 93.71 | 95.56 |
|  | CNN | 75.56 | 88.56 | 91.88 | 94.46 | 96.02 |
|  | GNN | 76.51 | 90.00 | 93.11 | 95.33 | 96.76 |

Fig. 6 depicts the loss curve of the proposed GNN model on 14-bus system during the training process. It can be observed that the proposed GNN model performs well in minimizing the MSE loss. The training loss decreases rapidly and then reaches the horizontal area after around 800 epochs. Table I shows the prediction accuracy of different neural network models. Deep neural network (DNN) and convolutional neural network (CNN) are used ad benchmarks [1]. The proposed GNN model has the highest prediction accuracy comparing the other models. For 14-bus system, the prediction accuracy is 96.66% with a 5% tolerance. Further investigations on 24-bus system revealed that GNN model maintains the highest prediction accuracy with different error tolerance.

Table II Comparison of Different Models

| System | Model | $R^2$ | MAX-E | MED-E | MAPE |
|---|---|---|---|---|---|
| 14 Bus | DNN | 0.9946 | 0.53 | 0.0279 | 0.0032 |
|  | CNN | 0.9963 | 0.28 | 0.0262 | 0.0031 |
|  | GNN | 0.9993 | 0.25 | 0.0183 | 0.0025 |
| 24 Bus | DNN | 0.9714 | 4.80 | 0.2154 | 0.0059 |
|  | CNN | 0.9733 | 4.55 | 0.1998 | 0.0055 |
|  | GNN | 0.9824 | 4.30 | 0.1855 | 0.0053 |

Table II summarizes the statics for all model architectures. The first column includes the R2 score results of all models on 14-bus and 24-bus system. It can be observed that the GNN model has the highest R2 score in both scenarios which is close to 1, implying that the predictions of GNN model approximate the real data points pretty well comparing to the other machine learning algorithm power flow models. The GNN model also has the lowest MAX-E, MED-E and MAPE statistical values.

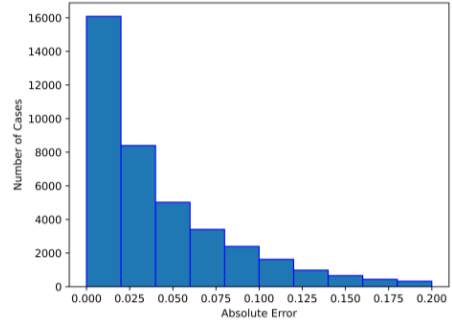

Fig. 7. Absolute error distribution of DNN neural network for 14-Bus System.

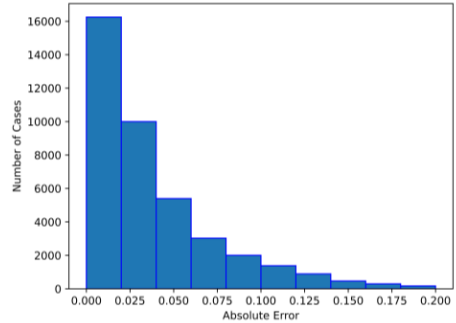

Fig. 8. Absolute error distribution of CNN neural network for 14-Bus System.

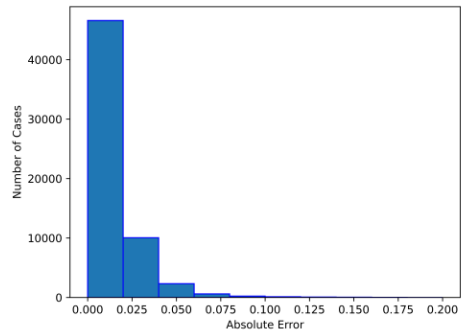

Fig. 9. Absolute error distribution of GNN neural network for 14-Bus System.

Figs. 7 - 9 compare the error distributions of GNN and benchmark models. Combining Table II and Fig. 7 – Fig. 9, it is observed that the proposed GNN model outperforms all other benchmarks. Over 95% of power flow prediction errors located within range 0.00-0.05 (MW), implying the predicted with the proposed GNN model are much closer to the reference values. Percentage relative difference (PRD) is calculated using the relative difference formula shown below,

$$PRD = \frac{2|y_i - \tilde{y}_i|}{|y_i + \tilde{y}_i|} \times 100\%. \tag{9}$$

Table III presents the PRD of line power flow. Results show that the accuracy of DCOF model is relatively lower than GNN and DNN models. For active power flow on 24-bus system, the mean PRD of GNN model is only 0.01 which is lower than DNN model at 0.02. Due to approximation error introduced in DCPF model, the maximal PRD of DCPF in both scenarios is much higher than data-driven models such as DNN and GNN models. Comparing to two other models, the proposed GNN model shows much better performance in predicting active power flow on both test systems. By comparing the standard deviations and median PRD of all models, we can also find that the proposed GNN model outperforms the traditional DCPF model.

Table III
PRD of Active Power Flow [MW] Comparing GNN Model and DCPF Model

| System | Model | Mean | Max | Min | Median | Std.Dev. |
|---|---|---|---|---|---|---|
| 14-bus | DNN | 0.07 | 15.79 | 0.00 | 0.003 | 0.34 |
| 14-bus | GNN | 0.05 | 13.45 | 0.00 | 0.001 | 0.26 |
| 14-bus | DCPF | 0.96 | 17.28 | 0.00 | 0.23 | 0.46 |
| 24- bus | DNN | 0.02 | 29.73 | 0.00 | 0.016 | 0.28 |
| 24- bus | GNN | 0.01 | 24.40 | 0.00 | 0.005 | 0.22 |
| 24- bus | DCPF | 1.36 | 30.63 | 0.00 | 0.53 | 0.93 |

V. CONCLUSIONS

With the potential to shift the entire computational effort to offline training, machine learning assisted power flow has become an increasingly interesting research direction. Traditional DCPF model introduces approximation error during power flow calculation. The GNN model shows great potential in processing power system data with geographical information embedded. The proposed GNN model can be trained offline using historic data, and this enables rapid determination of both line active power flows. It should be noted that the architecture of GNN model needs to be carefully selected, and sufficient resources are also needed for training the GNN model.

Results on 14-bus and 24-bus test systems show that the GNN model achieves results that closely align with those obtained from the ACPF method. GNN model shows high efficiency in power flow prediction. Compared to the DNN and CNN model, the proposed GNN model presents higher accuracy. In addition, the proposed GNN model performs faster with much better results compared to the DCPF model. Ongoing research and exploration of GNN applications in power systems indicate the vast potential of GNN in advancing and enhancing existing power system technologies.